\documentclass{article}

\usepackage{arxiv}

\usepackage[utf8]{inputenc} % allow utf-8 input
\usepackage[T1]{fontenc}    % use 8-bit T1 fonts
\usepackage{hyperref}       % hyperlinks
\usepackage{url}            % simple URL typesetting
\usepackage{booktabs}       % professional-quality tables
\usepackage{amsfonts}       % blackboard math symbols
\usepackage{nicefrac}       % compact symbols for 1/2, etc.
\usepackage{microtype}      % microtypography
\usepackage{lipsum}		% Can be removed after putting your text content
\usepackage{graphicx}
\usepackage{natbib}
\usepackage{doi}
\usepackage{amsmath}

\title{GeHirNet: A Gender-Aware Hierarchical Model for Voice Pathology Classification}

\author{ 
    \href{https://orcid.org/0000-0002-8414-0515}{\includegraphics[scale=0.06]{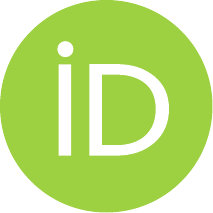}\hspace{1mm}Fan Wu} \\
	Centre for Digital Health Interventions\\
	ETH Zurich\\
	Zurich, Switzerland\\
	\texttt{fanwu@ethz.ch} \\
	\And
	\href{https://orcid.org/0009-0004-8087-744X}{\includegraphics[scale=0.06]{orcid.pdf}\hspace{1mm}Kaicheng Zhao} \\
	Institute of Mechanism Theory, Machine Dynamics and Robotics\\
	RWTH Aachen University\\
	Aachen, Germany\\
	\texttt{kaicheng.zhao@rwth-aachen.de} \\
	\And
	\href{https://orcid.org/0000-0002-4842-1117}{\includegraphics[scale=0.06]{orcid.pdf}\hspace{1mm}Elgar Fleisch} \\
	Centre for Digital Health Interventions \\
    ETH Zurich\\
	Zurich, Switzerland\\
    Centre for Digital Health Interventions \\
    University of St. Gallen\\
    St. Gallen, Switzerland \\
	\texttt{efleisch@ethz.ch} \\
	\And
	\href{https://orcid.org/0000-0002-3905-2380}{\includegraphics[scale=0.06]{orcid.pdf}\hspace{1mm}Filipe Barata} \\
	Centre for Digital Health Interventions\\
	ETH Zurich\\
	Zurich, Switzerland\\
	\texttt{fbarata@ethz.ch} \\
}

\hypersetup{
pdftitle={A template for the arxiv style},
pdfsubject={q-bio.NC, q-bio.QM},
pdfauthor={David S.~Hippocampus, Elias D.~Striatum},
pdfkeywords={First keyword, Second keyword, More},
}

\begin{document}
\maketitle

\begin{abstract}
	AI-based voice analysis shows promise for disease diagnostics, but existing classifiers often fail to accurately identify specific pathologies because of gender-related acoustic variations and the scarcity of data for rare diseases. We propose a novel two-stage framework that first identifies gender-specific pathological patterns using ResNet-50 on Mel spectrograms, then performs gender-conditioned disease classification. We address class imbalance through multi-scale resampling and time warping augmentation. Evaluated on a merged dataset from four public repositories, our two-stage architecture with time warping achieves state-of-the-art performance (97.63\% accuracy, 95.25\% MCC), with a 5\% MCC improvement over single-stage baseline. This work advances voice pathology classification while reducing gender bias through hierarchical modeling of vocal characteristics.

\end{abstract}

% keywords can be removed

\section{Introduction}

Voice pathologies, caused by factors such as infections, vocal fatigue, or neurological conditions such as muscular dystrophy, disrupt normal vocal fold vibration, resulting in strained, weak, or hoarse voices that degrade voice quality~\cite{titze1998principles}. Traditional diagnosis relies on invasive clinical procedures like laryngoscopy and stroboscopy, which require specialized equipment, cause patient discomfort, and incur high costs~\cite{sulica2013laryngoscopy}. 

Recent advances in signal processing and AI provide a promising non-invasive alternative: voice-based detection. By analyzing acoustic features from speech recordings, machine learning models have shown promise in distinguishing normal voices from pathological ones~\cite{al2020voice}. Recent studies reported 98\% accuracy with the MEEI database~\cite{amami2017incremental} and 95.41\% accuracy with the SVD dataset~\cite{mohammed2020voice}. 

While current approaches perform well in broad classifications (e.g., pathological vs. healthy), their accuracy declines when discriminating between specific pathology subtypes.
%For example, Muhammad et al. applied One-vs-Rest classification to differentiate between cysts, paralysis, and vocal polyp pathologies, achieving accuracies of 99.4\%, 93.2\%, and 91.5\%, respectively~\cite{muhammad2017voice}. 
One early study reported accuracies of only 67.8\% for male patients and 52.5\% for female patients when identifying five distinct laryngeal pathologies~\cite{muhammad2011automatic}. The diversity of diseases complicates voice pathology classification. Another study achieved 95\% accuracy in detecting laryngeal disorders but only 85\% for Parkinson's disease~\cite{orozco2015characterization}, further underscoring the difficulty in distinguishing between specific conditions.

These shortcomings are worsened by challenges such as gender disparities and data imbalance. Gender significantly influences voice characteristics, as male and female voices differ in pitch and frequency distribution~\cite{zimman2018transgender}. Recent studies have explored the gender-related patterns with deep learning models, such as convolutional neural network (CNN) in applications like emotion detection~\cite{dar2024exploring}. Narrowing the search space to a particular gender have enhanced classification accuracy~\cite{alnuaim2022speaker}. These advancements provide a novel perspective for leveraging gender-specific features in voice pathology classification. Researchers developed a cascaded model incorporating gender classification as a preliminary step, achieving 88.38\% accuracy for binary pathology detection~\cite{ksibi2023voice}. 

Another critical challenge is the class imbalance commonly present in voice pathology datasets, which has often been overlooked by researchers. High accuracy in such imbalanced datasets may not reliably reflect the robustness of a classifier, as models exhibit a bias toward the majority class. This issue has gained increasing attention, with researchers proposing various methods to address data imbalance~\cite{fan2021class}.

A recent review highlights the need for vowel and gender separation, while addressing the uneven distribution of rare pathologies~\cite{abdulmajeed2022review}. Accordingly, we hypothesize that the class imbalance commonly found in voice pathology datasets, often overlooked by researchers, is further aggravated by gender disparities. Differences in voice characteristics between males and females can result in unequal representation and introduce performance biases across various pathologies. 

To this end, we propose a novel hierarchical framework \textbf{GeHirNet} that first distinguishes male and female pathologies from healthy controls, then classifies specific diseases separately for each gender. This approach is the first to integrate gender-based differentiation in the initial stage of multi-class voice pathology classification. 

Our framework analyzes sustained vowel /a/ recordings from multiple datasets, including Coswara, SVD, ALS, and PC-GITA, to detect a diverse range of pathologies, including COVID-19, Parkinson’s Disease, Vocal Cord Paresis, Dysphonia, Laryngitis, and Amyotrophic Lateral Sclerosis (ALS).

The key contributions of this paper are as follows:
\begin{itemize}
    \item First, we propose a two-layer hierarchical architecture that integrates gender-specific patterns in the first stage, followed by pathology classification. 
    \item Second, we address class imbalance using two data augmentation techniques: multi-scale resampling and the novel application of time warping directly on the audio segment.
    \item Third, we conducted four experiments to validate our approach. We compared female and male pathology classifiers and interpreted the results from an audio feature perspective.
\end{itemize}

The code is available at \href{https://github.com/ADAMMA-CDHI-ETH-Zurich/GeHirNet}{GeHirNet's GitHub}. 

\section{Methods}

\subsection{Dataset}

We used four publicly available datasets. This study only considered vowel /a/ recordings from these datasets.

\textbf{Coswara (English)}~\cite{sharma2020coswara}: Designed for COVID-19 classification, %the Coswara dataset contains respiratory, cough, and speech sounds collected via worldwide crowdsourcing. 
it includes vowel /a/ recordings from 341 COVID-19 patients  (212 male, 129 female) and 924 healthy individuals (701 male, 223 female).

\textbf{ALS (Russian)}~\cite{vashkevich2019bulbar}: Designed for ALS diagnosis, a neurodegenerative disorder, it includes vowel /a/ recordings from 39 healthy individuals (22 male, 17 female) and 15 ALS patients (7 male, 8 female).

\textbf{PC-GITA (Spanish)}~\cite{orozco2014new}: Developed for Parkinson’s disease research, it includes 300 vowel /a/ recordings from 100 participants (each providing three samples), with 50 healthy controls (25 male, 25 female) and 50 Parkinson’s patients (25 male, 25 female).

\textbf{SVD (German)}~\cite{woldert2007saarbruecken}: This dataset includes over 70 voice disorders. We considered the most represented pathology categories: healthy controls (259 male, 428 female), Laryngitis (50 male, 32 female), Vocal cord paresis (70 male, 127 female), and Dysphonia (99 male, 174 female). In our work, the Dysphonia category merges seven subtypes from the SVD dataset: Dysphonia, Functional Dysphonia, Spasmodic Dysphonia, Psychogenic Dysphonia, Hypofunctional Dysphonia, Hypotonic Dysphonia, and Juvenile Dysphonia. 

We selected those datasets for robust gender-specific pathology analysis: (1) voice recordings from speakers of diverse language backgrounds to examine gender-based vocal patterns, (2) standardized sustained vowel /a/ segments to isolate fundamental voice features while controlling for linguistic variability, and (3) comprehensive pathology coverage with healthy controls to evaluate model generalizability. This minimizes population-specific biases while allowing systematic investigation of gender-dependent pathological signatures.

The distribution of original recordings in the merged dataset is summarized in Table~\ref{tab:data}. Health control recordings come from four datasets, while pathological recordings cover six diseases: COVID (Coswara), ALS (ALS), Parkinson's (PC-GITA), Laryngitis, Vocal Cord Paresis, and Dysphonia (SVD).

\subsection{Audio Pre-processing}

\subsubsection{Silence Removal}

Audio recordings in the datasets, particularly in Coswara, frequently contain silent periods. We followed established preprocessing practices in prior work~\cite{matias2022clinically} by implementing silence removal as our initial processing step. We first addressed inconsistencies by removing silent segments based on Root Mean Square (RMS) energy~\cite{sakhnov2009approach}. 
The audio signal was segmented into overlapping frames with a window size of $W = 2048$ with a hop size of $H = 512$. For each frame \( i \), the RMS energy \( E_{\text{RMS}}^{(i)} \) is computed as:  

\[
E_{\text{RMS}}^{(i)} = \sqrt{\frac{1}{W} \sum_{n=0}^{W-1} x_i^2[n]},
\]  

where \( x_i[n] \) represents the \( n \)-th sample in the \( i \)-th frame. Frames with RMS energy below an empirically determined threshold of \(10^{-3}\) were classified as silent. Since our data consists of only vowel recordings, we removed silent segments and concatenated voiced segments to reconstruct the audio. 

Direct concatenation, however, introduces artifacts, i.e., sharp transitions at segment boundaries. To prevent this, we applied linear crossfade, shaping fade profiles for smooth transitions and minimizing discontinuities~\cite{zavivska2022audio}. Linear crossfade interpolates between the overlapping regions of neighboring segments. Given a crossfade interval \( V \), the crossfaded output \( y[n] \) for each sample point \( n \in [0, V-1] \) is computed as:

\[
y[n] = (1 - \alpha_n) \cdot x_1[n] + \alpha_n \cdot x_2[n]
\]

where \( x_1[n] \) represents \( n \)-th sample from the last \( V \) samples of the previous segment, \( x_2[n] \) represents the \( n \)-th sample from the first \( V \) samples of the next segment. \( \alpha_n = \frac{n}{V - 1} \) is the the linear interpolation weight. We set the crossfade interval \( V \) as 512.

\subsubsection{Outlier Removal}

\begin{figure}[t]
  \centering
  \includegraphics[width=0.9\linewidth]{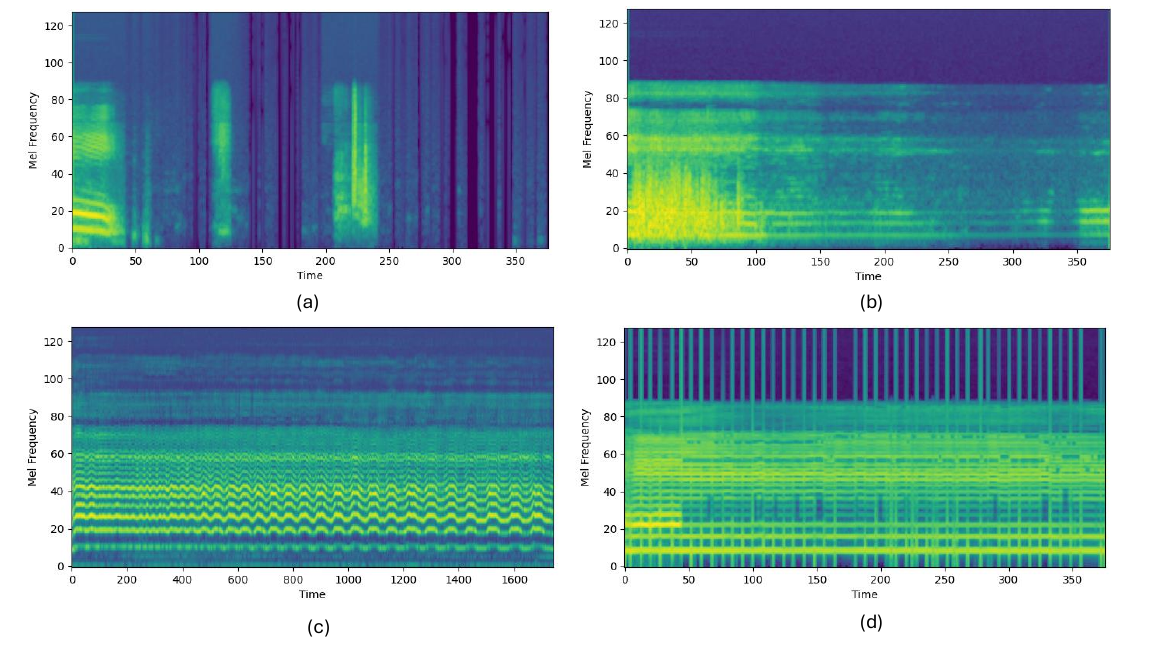}
  \caption{Example cases of outlier Mel spectrograms, (a) absence of valuable signal, (b) extreme transient shocks, (c) dominant global noise, and (d) erratic spiking signals.}
  \label{fig:outlier}
\end{figure}

After removing silence, we generated Mel spectrograms for all recordings using a window size of 2048 and a hop size of 512 for visual inspection~\cite{quan2022end}. As we expect sustained vowel recordings, we manually excluded recordings with abnormal spectral energy concentration, transient impulse artifacts, or aperiodic waveforms. Figure~\ref{fig:outlier} illustrates representative outlier cases identified in Mel spectrograms. 

\subsubsection{Normalization}

Min-Max normalization was applied to all valid recordings to compress the amplitude into the range [0, 1], to standardize the data across multiple datasets and to ensure consistency in amplitude levels. For an audio signal $x[n]$ of length $N$, the normalized signal $\tilde{x}[n]$ is computed as:

\[
\tilde{x}[n] = \frac{x[n] - \min(x)}{\max(x) - \min(x)}
\]

\subsubsection{Segmentation}
Finally we segmented audios into 1-second clips with a sliding window of 0.4 seconds~\cite{li2023rotating}. 

\subsection{Data Augmentation}

From the segment distribution, we observed a significant imbalance between healthy control and specific pathological diseases. To address this, we handled the data imbalance with data augmentation techniques.

\subsubsection{Multi-scale Resampling}

Data resampling is a common technique for mitigating data imbalance. In particular, we use sample-rate conversion to generate additional instances of minority classes and equalize their representation. In our dataset, ALS and PC-GITA have a sampling rate of 44.1 kHz, SVD uses 50 kHz, and Coswara includes both 48 kHz and 44.1 kHz rates. We applied sample-rate conversion to the classes with fewer samples by selecting a specific sampling rate from a range of 40 kHz to 50 kHz, with 125 Hz intervals. This process includes both upsampling and downsampling, with a sinc filter applied as an anti-aliasing filter. Resampling was repeated until the sample count for the underrepresented classes matched that of the most frequent class.

\subsubsection{Time Warping}

Time warping was originally applied to spectrograms by swapping two blocks of the same length from the time or frequency dimensions ~\cite{song2020specswap}. To compare fairly with resampling, we applied time warping directly to audio, which is plausible since vowel recordings are sustained and lack semantic information. We split each 1-second vowel recording into five segments and randomly shuffled their order before reassembly. To create smooth transitions between segments while preserving pathological characteristics, we applied crossfade with a short interval $V = 32$ during time warping augmentation.

\subsection{Mel Spectrogram}

Since Mel spectrograms effectively capture spectral information and speech variations~\cite{jegan2024pathological}, we converted all audio recordings into Mel spectrograms \( \{S_1, S_2, ..., S_N\} \). \( S_i(f,t) \) is the energy at frequency \( f \) and time \( t \) for the \( i \)-th sample, where \( f \) denotes the Mel filter banks (1 to 128), \( t \) denotes the time frame (1 to 98)~\cite{fan2022isnet}. For 50 kHz audio, we used a window length of 2048 and a hop length of 512. To ensure consistency across different sampling rates (44.1 kHz, 48 kHz, and resampled audio), we adjusted the short-time Fourier transform (STFT) parameters to generate consistent Mel spectrograms from 1-sec audio segments. This approach ensures consistent input size for feature extraction and model training, with slight variation in spectral resolution. As a result, the model received identical input representations, minimizing distortion and maintaining feature alignment.

\subsection{Backbone Model}

CNNs are highly effective in capturing spatial dependencies within two-dimensional features like Mel spectrograms and have proven useful for detecting subtle differences in pathological voices~\cite{wu2018convolutional}. Among CNN architectures, ResNet is particularly well-suited for pathological voice detection, achieving 98.13\% accuracy~\cite{jegan2024pathological}. The model ResNet-50 was pretrained on the ImageNet dataset~\cite{koonce2021resnet}, and we modified the input depth to 1 to match the single-channel Mel-spectrogram input. With Mel spectrograms as input representations, all layers were trained during model training.

\subsection{Two-Stage Hierarchical Architecture}

\begin{figure*}[t]
  \centering
  \includegraphics[width=\textwidth]{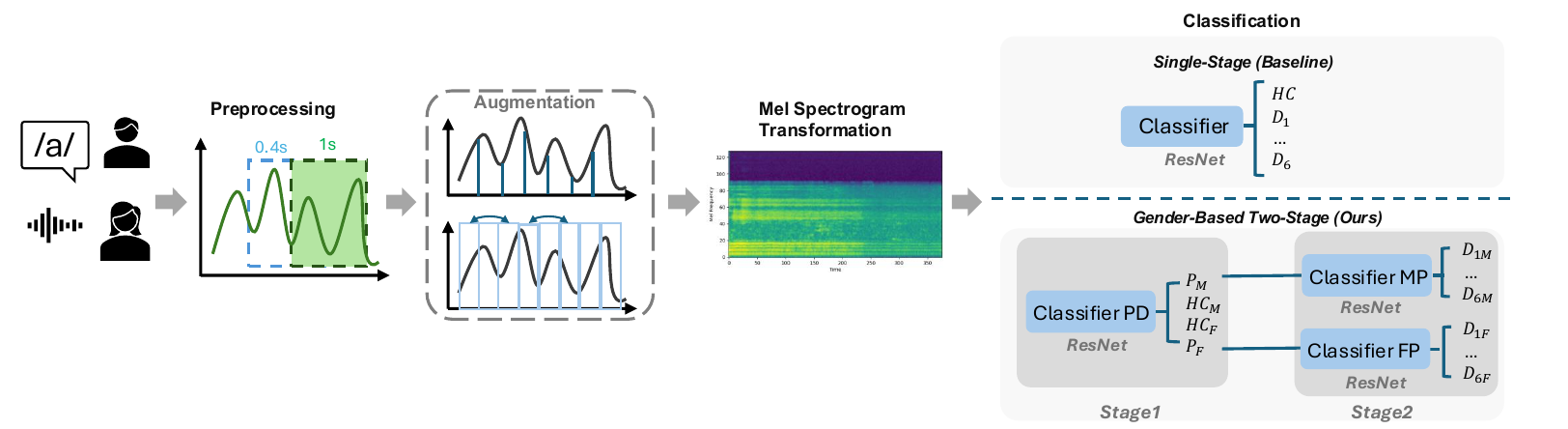} 
  \caption{Training Pipeline of Data Processing, Experiment Setup, and Model Architecture}
  \label{fig:pipeline}
\end{figure*}

Using ResNet as the backbone, we propose a two-stage hierarchical architecture \textbf{GeHirNet}, as illustrated in Figure~\ref{fig:pipeline}. The first stage focuses on gender-based \textit{Pathology Detection} (\textit{Classifier PD}), classifying samples into four target classes: male healthy control ($HC_M$), female healthy control ($HC_F$), male pathology ($P_M$), and female pathology ($P_F$). The male and female healthy control groups are later merged into a single healthy control category. 

In the second stage, we classify specific diseases separately within the male and female pathology groups. We achieve this by \textit{Pathology Classification}, \textit{Classifier FP} for female pathologies and \textit{Classifier MP} for male pathologies. \textit{Classifiers FP} and \textit{MP} distinguish among six diseases ($D_1$: COVID-19, $D_2$: Parkinson's, $D_3$: Dysphonia, $D_4$: Vocal Cord Paresis, $D_5$: laryngitis, $D_6$: ALS). The final disease classification result is obtained by merging the predictions from both gender-specific classifiers. All three classifiers, \textit{Classifiers PD}, \textit{MP}, and \textit{FP}, used the same ResNet-50 backbone and were trained separately.

\subsection{Experiment Setup}

To validate the effectiveness of the two-stage architecture and data augmentation techniques, we conducted four experiments independently. We split the dataset into a training set (80\%) and a test set (20\%) for evaluation. The split was stratified to ensure the same class distribution across training and testing data. All experiments used the same training and testing datasets.

\begin{itemize}
\item \textbf{Exp 1 (\textit{Baseline}) }: Single-stage 
%\item Exp 2.1: Balancing (Resampling) + One-stage architecture
%\item Exp 2.2: Balancing (Time warping) + One-stage architecture
\item \textbf{Exp 2 (\textit{GeHirNet}) }: Two-stage 
\item \textbf{Exp 3.1 (\textit{GeHirNet*}) }: Two-stage + Resampling
\item \textbf{Exp 3.2 (\textit{GeHirNet**}) }: Two-stage + Time warping

\end{itemize}
In \textbf{Exp 1} and \textbf{Exp 2}, no data augmentation was applied. After audio preprocessing, recordings were directly converted into Mel spectrograms as input images. \textbf{Exp 1} served as our \textbf{Baseline} and used a single-stage architecture to classify healthy controls ($HC$) and six specific pathologies ($D_1, ..., D_6$), while \textbf{Exp 2} employed two-stage hierarchical framework \textbf{GeHirNet}, where the gender-based \textit{Classifier PD} preceded separate classifiers for specific diseases (\textit{Classifiers MP and FP}). Both single-stage and two-stage architectures used ResNet-50 as the backbone. We compare \textbf{GeHirNet} with \textbf{Baseline} to evaluate the effectiveness of our two-stage hierarchical architecture.

We designed two experiments, \textbf{Exp 3.1} and \textbf{Exp 3.2}, to validate the effectiveness of data augmentation in addressing data imbalance. Data augmentation was applied only to the training set, keeping the test data unchanged. \textbf{Exp 3.1} and \textbf{Exp 3.2} incorporated resampling and time warping, respectively, for training data. After audio preprocessing, \textbf{Exp 3.1 (GeHirNet*)} applied resampling before converting audio to Mel spectrograms, while \textbf{Exp 3.2 (GeHirNet**)} applied time warping before conversion.  Since \textit{Classifiers PD}, \textit{MP}, and \textit{FP} were trained separately, augmentation was performed separately at two stages: \textit{Classifier PD} was balanced to match healthy control samples, while \textit{Classifiers FP} and \textit{MP} were balanced to the most frequent disease (COVID-19).

\subsection{Training \& Evaluation}

Taking Mel spectrograms as input, we trained the models using 5-fold cross-validation. We explored hyperparameter learning rates ($\eta$): $\{10^{-3}, 10^{-4}, 10^{-5}\}$, batch sizes ($B$): $\{32, 64\}$, training epochs ($E$): $\{10, 20, 30\}$. The Adam optimizer was used with default parameters ($\beta_1 = 0.9$, $\beta_2 = 0.999$) and no weight decay. For reproducibility, we fixed all random seeds to 42. Experiments were conducted on a single NVIDIA L4 GPU with PyTorch 2.6.0 and CUDA 12.4.

The optimization minimizes cross-entropy loss, which simultaneously maximizes the likelihood of correct class predictions and penalizes incorrect classifications. The labels of training data are encoded as one-hot vector $\mathbf{y}$.

\[
\mathbf{y} = [y_1, y_2, \ldots, y_C]
\]

The output dimension \( C \) of \(\mathbf{y}\) varies by experiment:

\begin{itemize}
    \item \textbf{Exp 1} (Single-stage): 
    \( C = 7 \) for six pathological diseases and health controls
    \item \textbf{Exp 2, 3.1 \& 3.2} (Two-stage):
    \begin{itemize}
            \item First stage (\textit{Pathology Detection}): \( C = 4 \) for male and female healthy control ($HC_M$, $HC_F$), male and female pathology ($P_M$, $P_F$).
        \item Second stage (\textit{Pathology Classification}): \( C = 6 \) for gender-specific pathologies, $D_{1M}, ..., D_{6M}$ for male and $D_{1F}, ..., D_{6F}$ for female.

    \end{itemize}
\end{itemize}

%The predicted probability distribution is denoted as $\hat{\mathbf{p}} = [\hat{p}_1, \hat{p}_2, \ldots, \hat{p}_C]$, whose components are computed by applying the softmax function to the output logits $\mathbf{z} = [z_1, z_2, \ldots, z_C]$:

%\[
%\hat{\mathbf{p}} = [\hat{p}_1, \hat{p}_2, \ldots, %\hat{p}_C]
%\]

%\[
%\hat{p}_i = \frac{e^{z_i}}{\sum_{j=1}^{C} e^{z_j}} %\quad \mathrm{for\ } i = 1, 2, \ldots, C
%\]

The predicted probability distribution is denoted as $\hat{\mathbf{p}} = [\hat{p}_1, \hat{p}_2, \ldots, \hat{p}_C]$ and the categorical cross-entropy loss for a single sample is defined as:
\[
\mathcal{L}^{(n)} = -\sum_{i=1}^{C} y_i^{(n)} \log(\hat{p}_i^{(n)})
\]

Averaging over the batch size \( B \), the total loss becomes:
\[
\mathcal{L} = \frac{1}{B} \sum_{n=1}^{B} \mathcal{L}^{(n)}
\]

The optimal hyperparameter was selected based on the average Matthews Correlation Coefficient (MCC) across validation folds, where $TP$, $FP$, $TN$ and $FN$ are true positives, false positives, true negatives and false negatives.

\[
\text{MCC} = \frac{TP \times TN - FP \times FN}{\sqrt{(TP+FP)(TP+FN)(TN+FP)(TN+FN)}}
\]

Using the best hyperparameters ($\eta$, $B$, $E^*$), we retrained the final model on the complete training set.

Finally, we assessed the model's performance on the unseen test data. We evaluated accuracy (proportion of correct predictions), weighted F1-score (harmonic mean of precision and recall, weighted by class frequency) and MCC (balanced metric considering all confusion matrix scores).

\[
\text{weighted F1} = \frac{\sum_{i=1}^{N} w_i \cdot \left(2 \cdot \frac{TP_i}{2TP_i + FP_i + FN_i}\right)}{\sum_{i=1}^{N} w_i}
\]

This metric combination was selected as accuracy provides an easily interpretable baseline, weighted F1-score specifically handles our dataset's class imbalance by weighting minority classes, and MCC serves as our primary robust metric since it reliably evaluates performance and remains invariant to class distribution, particularly crucial for our clinical diagnostic application with imbalanced data~\cite{chicco2020advantages}.

\subsection{Feature representation}

To explore gender disparities in audio features (i.e., Mel spectrograms) in pathology classification, we analyzed the similarity between \textit{Classifiers FP} and \textit{MP}. We calculated the Centered Kernel Alignment (CKA) score, which compares neural representations learned by different layers~\cite{cortes2012algorithms}. As \textit{Classifier FP} and \textit{MP} extracted spatial features with ResNet, we grouped ResNet into the convolution stem, four residual blocks, pooling, and fully connected layers, then calculated CKA scores for each group.

Given $n$ input samples, let $X \in \mathbb{R}^{n \times d_1}$ and $Y \in \mathbb{R}^{n \times d_2}$ denote feature matrices extracted from \textit{Classifiers MP} and \textit{FP}  respectively. The CKA similarity measure is computed as follows:

\[
\text{CKA}(X,Y) = \frac{\langle K_c, L_c \rangle_F}{\|K_c\|_F \|L_c\|_F}
\]
where:
\begin{itemize} 
\item $\tilde{X} = (I_n - \tfrac{1}{n}\mathbf{1}\mathbf{1}^\top)X$ (centered features)
\item $K_c = HKH^\top$ with $H = I_n - \tfrac{1}{n}\mathbf{1}\mathbf{1}^\top$ (double center)
\item $K = \tilde{X}\tilde{X}^\top$, $L = \tilde{Y}\tilde{Y}^\top$ (Gram matrices)
\item $\langle A,B \rangle_F = \mathrm{tr}(A^\top B)$ (Frobenius inner product)
\end{itemize}

\subsection{Statistical Analysis}

For each subgroup (e.g., gender × disease), we computed the mean Mel spectrogram by averaging the time-frequency representations (across all samples in the subgroup).
 
\[
\bar{S}(f,t) = \frac{1}{N} \sum_{i=1}^{N} S_i(f,t)
\]

Next, for each participant, we computed the mean power (in dB) of the Mel spectrogram across all time-frequency bins. We then derived group-level statistics by calculating the mean and standard deviation of these mean power values across all participants. 

\[
\text{Mean power (dB)} = \frac{1}{N \times M} \sum_{i=1}^{N} \sum_{j=1}^{M} 10 \cdot \log_{10}(S_{ij})
\]  
where \( N \) and \( M \) are numbers of frequency bins and time frames.

To compare power between genders for each disease, we performed a t-test when both groups followed a normal distribution. When normality assumptions were not met, we applied the Mann-Whitney U test as a non-parametric alternative.

\section{Results}

\subsection{Data Distribution}

During the preprocesssing, we applied silence removal and crossfade to 121 samples, 7 from PC-GITA and 114 from Coswara. After outlier removal, the final dataset comprised 1675 healthy recordings and 987 pathology recordings. Removal rates were balanced across subgroups: 6.33\% for female healthy, 5.14\% for female pathology, 7.38\% for male healthy, and 8.38\% for male pathology. The complete distribution is summarized in Table~\ref{tab:data}.

After the segmentation, the dataset contained 9250 (68.5\%) audio segments for healthy control, 1877 (13.89\%) for COVID-19, 1075 (7.96\%) for Parkinson’s, 597 (4.42\%) for Dysphonia, 412 (3.05\%) for Vocal Cord Paresis, 171 (1.27\%) for Laryngitis, 127 (0.94\%) for ALS.

\begin{table}[t]
\centering
%\resizebox{.95\columnwidth}{!}{
  \begin{tabular}{llll}
    \toprule
    \textbf{} & \textbf{Total} & \textbf{Female} & \textbf{Male} \\
    \midrule
    \textbf{$HC$}   & 1800 ({1675}) & 743 ({696}) & 1057 ({979}) \\
    \midrule
    Pathology   & 1058 ({987}) & 545 ({517}) & 513 ({470)} \\
    \hspace{0.2cm} $D_1$   & 341 ({283}) & 129 ({109}) & 212 ({174}) \\
    \hspace{0.2cm} $D_2$   & 150 ({149}) & 75 & 75 ({74}) \\
    \hspace{0.2cm} $D_3$  & 273 ({267}) & 174 ({170}) & 99 ({97}) \\
    \hspace{0.2cm} $D_4$   & 197 ({191}) & 127 ({123}) & 70 ({68}) \\
    \hspace{0.2cm} $D_5$    & 82 & 32 & 50 \\
    \hspace{0.2cm} $D_6$   & 15 & 8 & 7 \\
    \bottomrule
  \end{tabular}
\caption{Class distribution of the merged dataset. Number of original recordings and post-outlier removal recordings (in parentheses). Unparenthesized ones had no removals. $HC$: Health Control, $D_1$: COVID-19, $D_2$: Parkinson's, $D_3$: Dysphonia, $D_4$: Vocal Cord Paresis, $D_5$: Laryngitis, $D_6$: ALS. }
\label{tab:data}
\end{table}

\subsection{Comparison with Baseline}

Table~\ref{tab:results} illustrates the optimal hyperparameters (learning rate, batch size, and epochs) for each experiment, along with the evaluation metrics. For the two-stage architecture (\textbf{GeHirNet} and its augmented version), since the three classifiers (\textit{Classifier PD}, \textit{MP}, \textit{FP}) were trained separately, their hyperparameters are reported individually in the table. 

The two-stage architecture (\textbf{GeHirNet}) outperformed the single-stage model (\textbf{Baseline}), with a 3\% improvement in MCC (0.9363 vs. 0.9041), 1\% improvement in accuracy (0.9678 vs. 0.9526), and 1\% improvement in F1 score (0.9679 vs. 0.9513). After incorporating data augmentation, the two-stage model with resampling (\textbf{GeHirNet*}), achieved MCC of 0.9339, accuracy of 0.9671, and F1 of 0.9666, and performed similarly to \textbf{GeHirNet}. The two-stage model with time warping (\textbf{GeHirNet**}) achieved the highest performance with MCC of 0.9525, accuracy of 0.9763, and F1 of 0.9761. 

Taking MCC as the prioritized metric, \textbf{GeHirNet**}  achieved the best performance, improving MCC by 5\% over \textbf{Baseline}. This highlights the effectiveness of the two-stage hierarchical architecture in enhancing classification, with time warping as a data augmentation technique providing additional gains.

\begin{table*}[ht]
  \centering
  \begin{tabular}{lllll}
    \toprule
    \textbf{Exp} & \textbf{Optimal hyperparameters} & \textbf{Accuracy} & \textbf{F1} & \textbf{MCC} \\
    \midrule
    \textbf{Exp1 (Baseline)} & $10^{-3}$, 32, 30   & 0.9526 & 0.9513 &	0.9041 \\
    %\textbf{Exp2.1} ($10^{-3}$, 32, 30)   & 0.9560 &	0.9553 &	0.9116 \\
    %\textbf{Exp2.2} ($10^{-4}$, 32, 20)  & 0.9734 & 	0.9731 &	0.9468 \\
    \textbf{Exp2 (GeHirNet)} & [$10^{-4}$, 64, 30], [$10^{-3}$, 32, 30], [$10^{-4}$, 32, 20]  & 0.9678 & 0.9679 &	0.9363 \\
    \textbf{Exp3.1 (GeHirNet*)} & [$10^{-4}$, 64, 20], [$10^{-3}$, 32, 30],  [$10^{-3}$, 64, 30]   & 0.9671 & 	0.9666 &	0.9339 \\
    \textbf{Exp3.2 (GeHirNet**)} & [$10^{-4}$, 64, 30], [$10^{-3}$, 64, 20],  [$10^{-3}$, 64, 30]    & 0.9763 &  	0.9761 &	0.9525 \\
    \bottomrule
  \end{tabular}
  \caption{Experiment Results}
  \label{tab:results}
\end{table*}

\subsection{Comparison with Prior Work}

Our framework achieves better classification performance across pathologies compared to gender-agnostic models. While recent work reported 89.47\% accuracy for four-class discrimination (healthy vs. cyst, polyp, and paralysis)~\cite{al2021voice}, our gender-aware architecture demonstrates competitive performance, achieving an accuracy of 97.63\%, F1 of 97.61\% and MCC of 95.25\%. 

To address concerns regarding potential discrepancies in disease classification, particularly between functional, organic pathologies and neurological disorders, we evaluated the performance of our model on a disease-specific basis.
Table~\ref{tab:comparison} compares the accuracy of our framework with prior state-of-the-art (SOTA) methods, evaluated on the same dataset. We benchmark three variants: single-stage \textbf{Baseline}, two-stage architecture \textbf{GeHirNet}, and its enhanced variant with time warping, \textbf{GeHirNet**}. Results demonstrate that \textbf{GeHirNet} matches or exceeds SOTA performance across all six target diseases, with \textbf{GeHirNet**} achieving the highest overall accuracy.  

\begin{table*}[ht]
\centering
%\resizebox{.95\columnwidth}{!}{
  \begin{tabular}{lllll}
    \toprule
    \textbf{Disease} & \textbf{Prior SOTA} & \textbf{Baseline} & \textbf{GeHirNet} & \textbf{GeHirNet**}  \\
    \midrule
    Dysphonia    & 0.99~\cite{hammami2020voice}  &  0.986 &  0.990  & 0.992 \\
    Laryngitis   & 0.983~\cite{geng2025pathological}  & 0.997 & 0.997 & 0.998 \\
    Vocal cord paralysis (paresis)   & 0.941~\cite{hegde2025novel}  & 0.989  & 0.994 & 0.996 \\
    COVID-19    & 0.999~\cite{gidaye2025speech}  &  0.979  & 0.986 & 0.991 \\
    Parkinson’s Disease  & 0.997~\cite{zahid2020spectrogram}  &  0.995  & 0.997 & 0.997 \\
    ALS &  0.997~\cite{vashkevich2021classification}  &  1 & 1 & 1 \\
    \bottomrule
  \end{tabular}
\caption{Comparison of disease-specific accuracy between our models and prior SOTA methods, evaluated on the same dataset.}
\label{tab:comparison}
\end{table*}

Beyond disease-specific evaluations, we evaluated detection accuracy of healthy controls across all six datasets. \textbf{GeHirNet} achieved an accuracy of 97.6\% in detecting health controls, while \textbf{GeHirNet**} improved this to 99.3\%. This highlights the model's ability to generalize and detect health control across different datasets, rather than overfitting to specific data. It suggests that our model learns underlying pathological patterns, enabling cross-dataset generalization and mitigating issues like shortcut learning.

\subsection{Feature Representation}

\begin{figure}[t]
  \centering
  \includegraphics[width=0.9\linewidth]{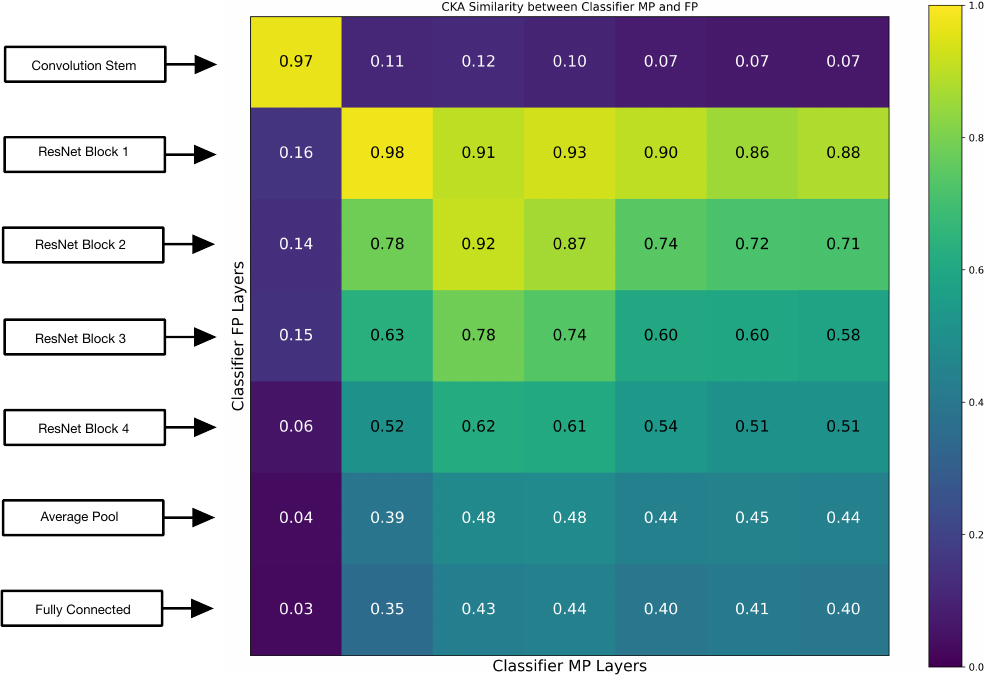}
  \caption{CKA scores between Classifier FP and MP across ResNet layers: convolution stem, four residual blocks, pooling, and fully connected.}
  \label{fig:cka}
\end{figure}

We analyzed gender-based differences in the learned features for pathology classification using CKA scores. As shown in Figure~\ref{fig:cka}, CKA scores reveal differences in neural representations learned by \textit{Classifiers MP} and \textit{FP}. We observe high CKA scores in the shallow layers and low CKA scores in the deep layers, indicating that the neural representations extracted are similar in the shallow layers but differ substantially in the deep layers. This suggests that the shallow layers tend to extract more universal and generic vowel features consistent across male and female speakers, while the deep layers capture more abstract, gender-specific, pathology-related spectral differences.

The shallow layers of ResNet encode basic spectro-temporal patterns, which remain consistent across genders, i.e., spectral envelopes~\cite{palaz2015analysis} and the F1/F2 ratios ~\cite{hillenbrand1995acoustic}. In contrast, deeper layers capture more gender-specific features, i.e., F0 and formants, with lower values in males and higher in females~\cite{gelfer2013speaking}, as well as glottal source features due to males' longer, higher-mass, and lower-tension vocal folds~\cite{munoz2013relevance}. It highlights the importance of considering gender-specific features in voice pathology classification. 

\subsection{Statistical Analysis}

\begin{figure*}[t]
  \centering
  \includegraphics[width=\textwidth]{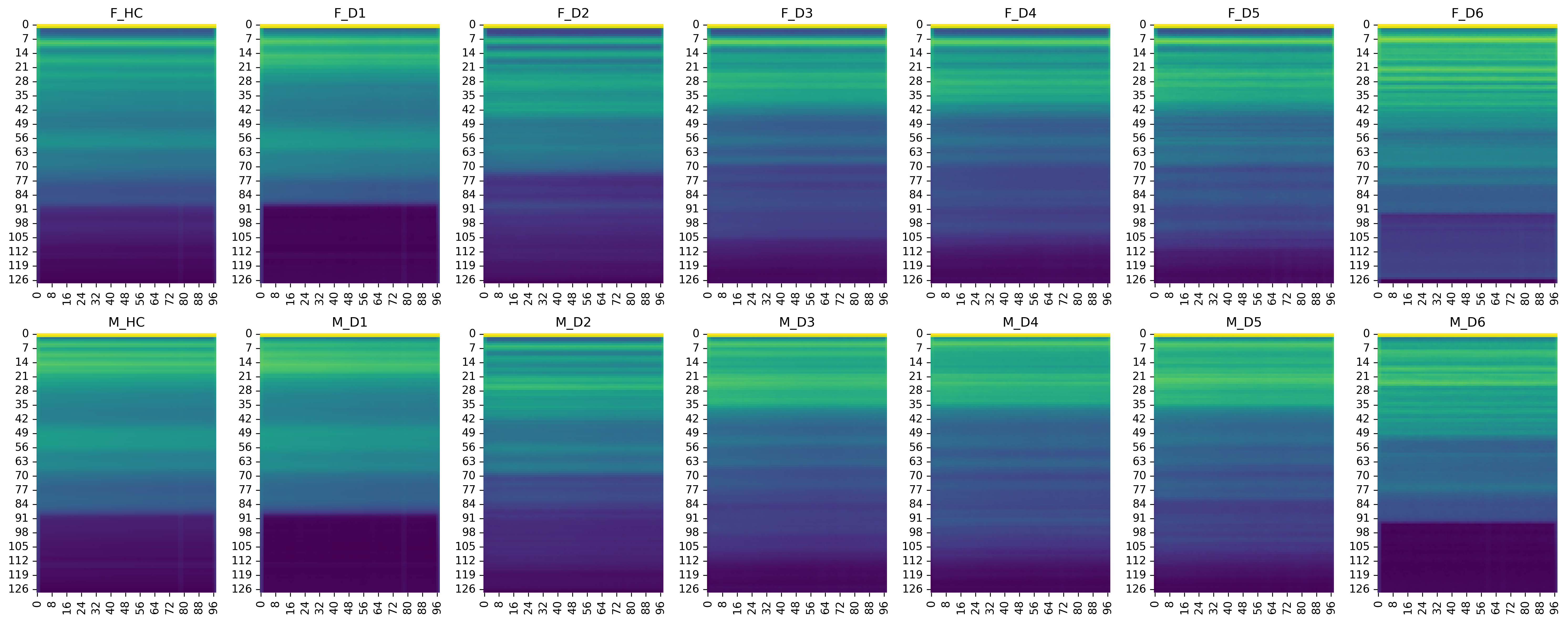} 
  \caption{Average Mel spectrograms of male and females for health controls and various diseases. $HC$: Health Control, $D_1$: COVID-19, $D_2$: Parkinson's, $D_3$: Dysphonia, $D_4$: Vocal Cord Paresis, $D_5$: Laryngitis, $D_6$: ALS. $F$: Female. $M$: Male.  }
  \label{fig:gender_mel}
\end{figure*}

As shown in Figure~\ref{fig:gender_mel}, we observe disease-specific acoustic patterns manifest differently between genders in Mel spectrograms. As shown in Table~\ref{tab:stat}, by computing group-level statistics of Mel spectrogram, we observed significant gender differences in Mel spectrogram power. Healthy females exhibited significantly higher power values than males ($\Delta$ = +4.92 dB), with this pattern persisting in ALS ($\Delta$  = +4.393 dB). Females in COVID-19 ($\Delta$ = -1.523 dB) and Vocal Cord Paresis ($\Delta$  = -0.919 dB) showed lower power values. The gender disparity diminished in Parkinson's, Dysphonia and Laryngitis, where no significant difference was detected. The identified gender differences in pathological Mel spectrograms validates the rationale behind our gender-aware model. By explicitly modeling these differences, our approach outperforms gender-agnostic systems, achieving more precise pathology detection.

\begin{table}[ht]
\centering
%\resizebox{.95\columnwidth}{!}{
  \begin{tabular}{lllll}
    \toprule
    \textbf{Disease} & \textbf{Female (dB)} & \textbf{Male (dB)} & \textbf{  $\Delta$ (F-M)} \\
    \midrule
    $HC$  & $-9.63 \pm 9.49$  &  $-14.55 \pm 9.16$ &  $+4.921*$ \\
    $D_1$   & $-22.38 \pm 7.22$  & $-20.86 \pm 7.50$  & $-1.523*$ \\
    $D_2$  & $-2.80 \pm 3.74$  & $-2.99 \pm 3.14$  & $+0.192$ \\
    $D_3$   & $-2.69 \pm 2.49$  &  $-2.96 \pm 2.33$  & $+0.267$ \\
    $D_4$  & $-2.56 \pm 3.42$  &  $-1.64 \pm 4.02$  & $-0.919*$ \\
    $D_5$ &  $-3.05 \pm 3.04$  &  $-2.43 \pm 3.42$ & $-0.618$ \\
    $D_6$ &  $-5.23 \pm 7.63$  &  $-9.62 \pm 4.55$ & $+4.393*$ \\

    \bottomrule
  \end{tabular}
\caption{Gender differences in Mel spectrogram power across diseases. $HC$: Health Control, $D_1$: COVID-19, $D_2$: Parkinson's, $D_3$: Dysphonia, $D_4$: Vocal Cord Paresis, $D_5$: Laryngitis, $D_6$: ALS. * indicate statistically significant differences ($p<0.05$). }
\label{tab:stat}
\end{table}

\subsection{Related Work}

Prior research on hierarchical voice pathology classification, such as Cordeiro's three-class system achieving 83\% accuracy~\cite{cordeiro2017hierarchical}, has not accounted for gender differences. Our work advances this paradigm by introducing gender-based hierarchical classifiers that explicitly address class imbalance—a critical issue exacerbated by gender disparities.  

Unlike previous hierarchical frameworks that used disconnected modular blocks (e.g., separating gender detection from pathology classification)~\cite{ksibi2023voice}, our two-layer model integrated gender representations directly into multi-class pathology detection. We implemented an end-to-end framework where the first layer outputs four gender-health classes, enabling disease classification in the second layer based on gender-aware representations.  

Rather than treating gender and pathology as separate, sequential tasks, we explicitly embedded gender into the learning process, enabling the model to capture distinct, gender-dependent pathological features. By integrating gender directly into the classification pipeline, our approach mitigates performance disparities caused by gender-based variation.  

\section{Conclusion}

%We acknowledge the limitations of our work, particularly that we only considered vowel /a/ recordings. For more complex recordings, such as full speech, time-warping may no longer be suitable. Additionally, given the use of multiple datasets, it is challenging to determine whether the features we captured are influenced by variations in the recording environment. 

By introducing a two-stage hierarchical architecture that first detects pathology separately by gender and then classifies specific diseases, our work addresses critical limitations of traditional single-stage approaches. Conventional methods often struggled with severe class imbalance, particularly between healthy controls and rare diseases, and overlooked gender-dependent variations in voice data. Our framework mitigates these issues by (1) separating healthy and pathological samples in the first stage to reduce imbalance, and (2) explicitly modeling gender-specific features for more robust disease classification.  

Through controlled experiments, we demonstrate that our two-layer design alone outperforms prior approaches, even without augmentations. Further gains are achieved with targeted techniques like time warping, which enhances accuracy for underrepresented classes. By systematically addressing both class imbalance and gender-based variability, our model enhances diagnostic accuracy and reliability in data-limited scenarios. This advancement paves the way for scalable and cost-effective voice pathology screening.

\bibliographystyle{unsrtnat}
\bibliography{references}  %%% Uncomment this line and comment out the ``thebibliography'' section below to use the external .bib file (using bibtex) .

\begin{thebibliography}{41}
\providecommand{\natexlab}[1]{#1}
\providecommand{\url}[1]{\texttt{#1}}
\expandafter\ifx\csname urlstyle\endcsname\relax
  \providecommand{\doi}[1]{doi: #1}\else
  \providecommand{\doi}{doi: \begingroup \urlstyle{rm}\Url}\fi

\bibitem[Titze and Martin(1998)]{titze1998principles}
Ingo~R Titze and Daniel~W Martin.
\newblock Principles of voice production, 1998.

\bibitem[Sulica(2013)]{sulica2013laryngoscopy}
Lucian Sulica.
\newblock Laryngoscopy, stroboscopy and other tools for the evaluation of voice disorders.
\newblock \emph{Otolaryngol Clin North Am}, 46\penalty0 (1):\penalty0 21--30, 2013.

\bibitem[AL-Dhief et~al.(2020)AL-Dhief, Latiff, Malik, Sabri, Baki, Albadr, Abbas, Hussein, and Mohammed]{al2020voice}
Fahad~Taha AL-Dhief, Nurul Mu’azzah~Abdul Latiff, Nik Noordini Nik~Abd Malik, Naseer Sabri, Marina~Mat Baki, Musatafa Abbas~Abbood Albadr, Aymen~Fadhil Abbas, Yaqdhan~Mahmood Hussein, and Mazin~Abed Mohammed.
\newblock Voice pathology detection using machine learning technique.
\newblock In \emph{2020 IEEE 5th international symposium on telecommunication technologies (ISTT)}, pages 99--104. IEEE, 2020.

\bibitem[Amami and Smiti(2017)]{amami2017incremental}
Rimah Amami and Abir Smiti.
\newblock An incremental method combining density clustering and support vector machines for voice pathology detection.
\newblock \emph{Computers \& Electrical Engineering}, 57:\penalty0 257--265, 2017.

\bibitem[Mohammed et~al.(2020)Mohammed, Abdulkareem, Mostafa, Khanapi Abd~Ghani, Maashi, Garcia-Zapirain, Oleagordia, Alhakami, and Al-Dhief]{mohammed2020voice}
Mazin~Abed Mohammed, Karrar~Hameed Abdulkareem, Salama~A Mostafa, Mohd Khanapi Abd~Ghani, Mashael~S Maashi, Begonya Garcia-Zapirain, Ibon Oleagordia, Hosam Alhakami, and Fahad~Taha Al-Dhief.
\newblock Voice pathology detection and classification using convolutional neural network model.
\newblock \emph{Applied Sciences}, 10\penalty0 (11):\penalty0 3723, 2020.

\bibitem[Muhammad et~al.(2011)Muhammad, Alsulaiman, Mahmood, and Ali]{muhammad2011automatic}
Ghulam Muhammad, Mansour Alsulaiman, Awais Mahmood, and Zulfiqar Ali.
\newblock Automatic voice disorder classification using vowel formants.
\newblock In \emph{2011 IEEE international conference on multimedia and expo}, pages 1--6. IEEE, 2011.

\bibitem[Orozco-Arroyave et~al.(2015)Orozco-Arroyave, Belalcazar-Bolanos, Arias-Londo{\~n}o, Vargas-Bonilla, Skodda, Rusz, Daqrouq, H{\"o}nig, and N{\"o}th]{orozco2015characterization}
Juan~Rafael Orozco-Arroyave, Elkyn~Alexander Belalcazar-Bolanos, Juli{\'a}n~David Arias-Londo{\~n}o, Jes{\'u}s~Francisco Vargas-Bonilla, Sabine Skodda, Jan Rusz, Khaled Daqrouq, Florian H{\"o}nig, and Elmar N{\"o}th.
\newblock Characterization methods for the detection of multiple voice disorders: neurological, functional, and laryngeal diseases.
\newblock \emph{IEEE journal of biomedical and health informatics}, 19\penalty0 (6):\penalty0 1820--1828, 2015.

\bibitem[Zimman(2018)]{zimman2018transgender}
Lal Zimman.
\newblock Transgender voices: Insights on identity, embodiment, and the gender of the voice.
\newblock \emph{Language and Linguistics Compass}, 12\penalty0 (8):\penalty0 e12284, 2018.

\bibitem[Dar and Delhibabu(2024)]{dar2024exploring}
Gh~Mohmad Dar and Radhakrishnan Delhibabu.
\newblock Exploring emotion detection in kashmiri audio reviews using the fusion model of cnn, lstm, and rnn: gender-specific speech patterns and performance analysis.
\newblock \emph{International Journal of Information Technology}, pages 1--19, 2024.

\bibitem[Alnuaim et~al.(2022)Alnuaim, Zakariah, Shashidhar, Hatamleh, Tarazi, Shukla, and Ratna]{alnuaim2022speaker}
Abeer~Ali Alnuaim, Mohammed Zakariah, Chitra Shashidhar, Wesam~Atef Hatamleh, Hussam Tarazi, Prashant~Kumar Shukla, and Rajnish Ratna.
\newblock Speaker gender recognition based on deep neural networks and resnet50.
\newblock \emph{Wireless Communications and Mobile Computing}, 2022\penalty0 (1):\penalty0 4444388, 2022.

\bibitem[Ksibi et~al.(2023)Ksibi, Hakami, Alturki, Asiri, Zakariah, and Ayadi]{ksibi2023voice}
Amel Ksibi, Nada~Ali Hakami, Nazik Alturki, Mashael~M Asiri, Mohammed Zakariah, and Manel Ayadi.
\newblock Voice pathology detection using a two-level classifier based on combined cnn--rnn architecture.
\newblock \emph{Sustainability}, 15\penalty0 (4):\penalty0 3204, 2023.

\bibitem[Fan et~al.(2021)Fan, Wu, Zhou, Zhang, and Tao]{fan2021class}
Ziqi Fan, Yuanbo Wu, Changwei Zhou, Xiaojun Zhang, and Zhi Tao.
\newblock Class-imbalanced voice pathology detection and classification using fuzzy cluster oversampling method.
\newblock \emph{Applied Sciences}, 11\penalty0 (8):\penalty0 3450, 2021.

\bibitem[Abdulmajeed et~al.(2022)Abdulmajeed, Al-Khateeb, and Mohammed]{abdulmajeed2022review}
Nuha~Qais Abdulmajeed, Belal Al-Khateeb, and Mazin~Abed Mohammed.
\newblock A review on voice pathology: Taxonomy, diagnosis, medical procedures and detection techniques, open challenges, limitations, and recommendations for future directions.
\newblock \emph{Journal of Intelligent Systems}, 31\penalty0 (1):\penalty0 855--875, 2022.

\bibitem[Sharma et~al.(2020)Sharma, Krishnan, Kumar, Ramoji, Chetupalli, Ghosh, Ganapathy, et~al.]{sharma2020coswara}
Neeraj Sharma, Prashant Krishnan, Rohit Kumar, Shreyas Ramoji, Srikanth~Raj Chetupalli, Prasanta~Kumar Ghosh, Sriram Ganapathy, et~al.
\newblock Coswara--a database of breathing, cough, and voice sounds for covid-19 diagnosis.
\newblock \emph{arXiv preprint arXiv:2005.10548}, 2020.

\bibitem[Vashkevich et~al.(2019)Vashkevich, Petrovsky, and Rushkevich]{vashkevich2019bulbar}
Maxim Vashkevich, Alexander Petrovsky, and Yuliya Rushkevich.
\newblock Bulbar als detection based on analysis of voice perturbation and vibrato.
\newblock In \emph{2019 Signal Processing: Algorithms, Architectures, Arrangements, and Applications (SPA)}, pages 267--272. IEEE, 2019.

\bibitem[Orozco-Arroyave et~al.(2014)Orozco-Arroyave, Arias-Londo{\~n}o, Vargas-Bonilla, Gonzalez-R{\'a}tiva, and N{\"o}th]{orozco2014new}
Juan~Rafael Orozco-Arroyave, Juli{\'a}n~David Arias-Londo{\~n}o, Jes{\'u}s~Francisco Vargas-Bonilla, Mar{\'\i}a~Claudia Gonzalez-R{\'a}tiva, and Elmar N{\"o}th.
\newblock New spanish speech corpus database for the analysis of people suffering from parkinson's disease.
\newblock In \emph{Lrec}, pages 342--347, 2014.

\bibitem[Woldert-Jokisz(2007)]{woldert2007saarbruecken}
Bogdan Woldert-Jokisz.
\newblock Saarbruecken voice database.
\newblock 2007.

\bibitem[Matias et~al.(2022)Matias, Costa, Carreiro, Gamboa, Sousa, Gomez, Sousa, Neuparth, Carreiro-Martins, and Soares]{matias2022clinically}
Pedro Matias, Joao Costa, Andr{\'e}~V Carreiro, Hugo Gamboa, Ines Sousa, Pedro Gomez, Joana Sousa, Nuno Neuparth, Pedro Carreiro-Martins, and Filipe Soares.
\newblock Clinically relevant sound-based features in covid-19 identification: robustness assessment with a data-centric machine learning pipeline.
\newblock \emph{IEEE Access}, 10:\penalty0 105149--105168, 2022.

\bibitem[Sakhnov et~al.(2009)Sakhnov, Verteletskaya, and Simak]{sakhnov2009approach}
Kirill Sakhnov, Ekaterina Verteletskaya, and Boris Simak.
\newblock Approach for energy-based voice detector with adaptive scaling factor.
\newblock \emph{IAENG International Journal of Computer Science}, 36\penalty0 (4), 2009.

\bibitem[Z{\'a}vi{\v{s}}ka et~al.(2022)Z{\'a}vi{\v{s}}ka, Rajmic, and Mokr{\`y}]{zavivska2022audio}
Pavel Z{\'a}vi{\v{s}}ka, Pavel Rajmic, and Ond{\v{r}}ej Mokr{\`y}.
\newblock Audio declipping performance enhancement via crossfading.
\newblock \emph{Signal Processing}, 192:\penalty0 108365, 2022.

\bibitem[Quan et~al.(2022)Quan, Ren, Luo, Chen, and Ling]{quan2022end}
Changqin Quan, Kang Ren, Zhiwei Luo, Zhonglue Chen, and Yun Ling.
\newblock End-to-end deep learning approach for parkinson’s disease detection from speech signals.
\newblock \emph{Biocybernetics and Biomedical Engineering}, 42\penalty0 (2):\penalty0 556--574, 2022.

\bibitem[Li et~al.(2023)Li, Lu, Tang, Zhang, Tian, Cui, Jiang, Li, and Jiang]{li2023rotating}
Fan Li, Zixiao Lu, Junyue Tang, Weiwei Zhang, Yahui Tian, Zhongyu Cui, Fei Jiang, Honglang Li, and Shengyuan Jiang.
\newblock Rotating machinery state recognition based on mel-spectrum and transfer learning.
\newblock \emph{Aerospace}, 10\penalty0 (5):\penalty0 480, 2023.

\bibitem[Song et~al.(2020)Song, Wu, Huang, Su, and Meng]{song2020specswap}
Xingcheng Song, Zhiyong Wu, Yiheng Huang, Dan Su, and Helen Meng.
\newblock Specswap: A simple data augmentation method for end-to-end speech recognition.
\newblock In \emph{Interspeech}, pages 581--585, 2020.

\bibitem[Jegan and Jayagowri(2024)]{jegan2024pathological}
Roohum Jegan and R~Jayagowri.
\newblock Pathological voice detection using optimized deep residual neural network and explainable artificial intelligence.
\newblock \emph{Multimedia Tools and Applications}, pages 1--27, 2024.

\bibitem[Fan et~al.(2022)Fan, Xu, Cai, and Xing]{fan2022isnet}
Weiquan Fan, Xiangmin Xu, Bolun Cai, and Xiaofen Xing.
\newblock Isnet: Individual standardization network for speech emotion recognition.
\newblock \emph{IEEE/ACM Transactions on Audio, Speech, and Language Processing}, 30:\penalty0 1803--1814, 2022.

\bibitem[Wu et~al.(2018)Wu, Soraghan, Lowit, and Di~Caterina]{wu2018convolutional}
Huiyi Wu, John Soraghan, Anja Lowit, and Gaetano Di~Caterina.
\newblock Convolutional neural networks for pathological voice detection.
\newblock In \emph{2018 40th annual international conference of the ieee engineering in medicine and biology society (EMBC)}, pages 1--4. IEEE, 2018.

\bibitem[Koonce and Koonce(2021)]{koonce2021resnet}
Brett Koonce and Brett Koonce.
\newblock Resnet 50.
\newblock \emph{Convolutional neural networks with swift for tensorflow: image recognition and dataset categorization}, pages 63--72, 2021.

\bibitem[Chicco and Jurman(2020)]{chicco2020advantages}
Davide Chicco and Giuseppe Jurman.
\newblock The advantages of the matthews correlation coefficient (mcc) over f1 score and accuracy in binary classification evaluation.
\newblock \emph{BMC genomics}, 21:\penalty0 1--13, 2020.

\bibitem[Cortes et~al.(2012)Cortes, Mohri, and Rostamizadeh]{cortes2012algorithms}
Corinna Cortes, Mehryar Mohri, and Afshin Rostamizadeh.
\newblock Algorithms for learning kernels based on centered alignment.
\newblock \emph{The Journal of Machine Learning Research}, 13:\penalty0 795--828, 2012.

\bibitem[Al-Dhief et~al.(2021)Al-Dhief, Baki, Latiff, Malik, Salim, Albader, Mahyuddin, and Mohammed]{al2021voice}
Fahad~Taha Al-Dhief, Marina~Mat Baki, Nurul Mu’azzah~Abdul Latiff, Nik Noordini Nik~Abd Malik, Naseer~Sabri Salim, Musatafa Abbas~Abbood Albader, Nor~Muzlifah Mahyuddin, and Mazin~Abed Mohammed.
\newblock Voice pathology detection and classification by adopting online sequential extreme learning machine.
\newblock \emph{IEEE Access}, 9:\penalty0 77293--77306, 2021.

\bibitem[Hammami et~al.(2020)Hammami, Salhi, and Labidi]{hammami2020voice}
I~Hammami, L~Salhi, and S~Labidi.
\newblock Voice pathologies classification and detection using emd-dwt analysis based on higher order statistic features.
\newblock \emph{Irbm}, 41\penalty0 (3):\penalty0 161--171, 2020.

\bibitem[Geng et~al.(2025)Geng, Liang, Shan, Xiao, Wang, and Wei]{geng2025pathological}
Lei Geng, Yan Liang, Hongfeng Shan, Zhitao Xiao, Wei Wang, and Mei Wei.
\newblock Pathological voice detection and classification based on multimodal transmission network.
\newblock \emph{Journal of Voice}, 39\penalty0 (3):\penalty0 591--601, 2025.

\bibitem[Hegde et~al.(2025)Hegde, Shenoy, and Devaraja]{hegde2025novel}
K~Jayashree Hegde, K~Manjula Shenoy, and K~Devaraja.
\newblock A novel stacked model for classification of vocal cord paralysis over imbalanced vocal data.
\newblock \emph{IEEE Access}, 2025.

\bibitem[Gidaye et~al.(2025)Gidaye, Barage, Dighe, Ezzine, Turkar, and Nagare]{gidaye2025speech}
Girish Gidaye, Abhay Barage, Nirmayee Dighe, Kadria Ezzine, Varsha Turkar, and Gajanan Nagare.
\newblock Speech signals as biomarkers: using glottal features for non-invasive covid-19 testing.
\newblock \emph{International Journal of Biomedical Engineering and Technology}, 47\penalty0 (1):\penalty0 65--85, 2025.

\bibitem[Zahid et~al.(2020)Zahid, Maqsood, Durrani, Bakhtyar, Baber, Jamal, Mehmood, and Song]{zahid2020spectrogram}
Laiba Zahid, Muazzam Maqsood, Mehr~Yahya Durrani, Maheen Bakhtyar, Junaid Baber, Habibullah Jamal, Irfan Mehmood, and Oh-Young Song.
\newblock A spectrogram-based deep feature assisted computer-aided diagnostic system for parkinson’s disease.
\newblock \emph{IEEE Access}, 8:\penalty0 35482--35495, 2020.

\bibitem[Vashkevich and Rushkevich(2021)]{vashkevich2021classification}
Maxim Vashkevich and Yu~Rushkevich.
\newblock Classification of als patients based on acoustic analysis of sustained vowel phonations.
\newblock \emph{Biomedical Signal Processing and Control}, 65:\penalty0 102350, 2021.

\bibitem[Palaz et~al.(2015)Palaz, Collobert, et~al.]{palaz2015analysis}
Dimitri Palaz, Ronan Collobert, et~al.
\newblock Analysis of cnn-based speech recognition system using raw speech as input.
\newblock 2015.

\bibitem[Hillenbrand et~al.(1995)Hillenbrand, Getty, Clark, and Wheeler]{hillenbrand1995acoustic}
James Hillenbrand, Laura~A Getty, Michael~J Clark, and Kimberlee Wheeler.
\newblock Acoustic characteristics of american english vowels.
\newblock \emph{The Journal of the Acoustical society of America}, 97\penalty0 (5):\penalty0 3099--3111, 1995.

\bibitem[Gelfer and Bennett(2013)]{gelfer2013speaking}
Marylou~Pausewang Gelfer and Quinn~E Bennett.
\newblock Speaking fundamental frequency and vowel formant frequencies: Effects on perception of gender.
\newblock \emph{Journal of Voice}, 27\penalty0 (5):\penalty0 556--566, 2013.

\bibitem[Mu{\~n}oz~Mulas et~al.(2013)Mu{\~n}oz~Mulas, Mart{\'\i}nez~Olalla, G{\'o}mez~Vilda, {\'A}lvarez~Marquina, and Mazaira~Fern{\'a}ndez]{munoz2013relevance}
Cristina Mu{\~n}oz~Mulas, Rafael Mart{\'\i}nez~Olalla, Pedro G{\'o}mez~Vilda, Agust{\'\i}n {\'A}lvarez~Marquina, and Luis~Miguel Mazaira~Fern{\'a}ndez.
\newblock Relevance of the glottal pulse and the vocal tract in gender detection.
\newblock 2013.

\bibitem[Cordeiro et~al.(2017)Cordeiro, Fonseca, Guimar{\~a}es, and Meneses]{cordeiro2017hierarchical}
Hugo Cordeiro, Jos{\'e} Fonseca, Isabel Guimar{\~a}es, and Carlos Meneses.
\newblock Hierarchical classification and system combination for automatically identifying physiological and neuromuscular laryngeal pathologies.
\newblock \emph{Journal of voice}, 31\penalty0 (3):\penalty0 384--e9, 2017.

\end{thebibliography}





% \end{thebibliography}

\end{document}